\documentclass[pre,preprint,groupedaddress, showkeys,showpacs]{revtex4}

\usepackage{graphicx}
\usepackage{amsmath}
\usepackage{mathrsfs}

\begin{document}


\title{Entropic Ratchet transport of interacting active Browanian particles}
\author{Bao-quan  Ai $^{1}$} \email[Email: ]{aibq@hotmail.com}
\author{Ya-feng He$^{2}$}
 \author{Wei-rong Zhong$^{3}$} 
\affiliation{$^{1}$Laboratory of Quantum Engineering and Quantum Materials, School of Physics and Telecommunication
Engineering, South China Normal University, 510006 Guangzhou, China\\
$^{2}$College of Physics Science and Technology, Hebei University, 071002 Baoding, China.\\
$^{3}$Department of Physics and Siyuan Laboratory, College of Science and Engineering, Jinan University, 510632 Guangzhou, China.}


\date{\today}
\begin{abstract}
  \indent   Directed transport of interacting active (self-propelled)Brownian particles is numerically investigated in confined geometries (entropic barriers). The self-propelled velocity can break thermodynamical equilibrium and induce the directed transport.  It is found that the interaction between active particles can greatly affect the ratchet transport.  For attractive particles, on increasing the interaction strength, the average velocity firstly decreases to its minima, then increases, and finally decreases to zero.  For repulsive particles,  when the interaction  is very weak, there exists a critical interaction at which the average velocity is minimal, nearly tends to zero, however, for the strong interaction, the average velocity is independent of the interaction.
  \end{abstract}

\pacs{05. 40. -a, 05. 60. Cd,  82. 70. Dd}



\maketitle
\section {Introduction}
\indent Diffusion in confined geometries is ubiquitous in nature.
The reduction of the coordinates in confined structures can
cause the appearance of remarkable entropic effects. More recently,
physicists have started to study entropic effects in
out-of-equilibrium phenomena such as transport of particles in corrugated channels\cite{rmp}.
Based on the geometry of the channel wall,  corrugated channels fall into two categories: Compartmentalized
channe\cite{ Marchesoni, Makhnovskii, hanggi2} and smoothly
corrugated channels \cite{Hanggi1,
Zwanzig,Kalinay,Laachi,Mondal,Ai,Dagdug}. The relevance of entropic
barriers to promote entropic transport in confined environments has
been recognized in a variety of situations that include molecular
transport in zeolites, ionic channels, or in microfluidic devices.
The entropic transport in these systems yields important and exhibits peculiar properties.

\indent In previous works, the entropic transport mainly focused
on passive  particles and few works on the entropic transport have
involved active Brownian particles. However,  active
matters in biological and physical systems have been studied theoretically and experimentally \cite{ai2,Schimansky, tailleur,fily,
kaiser, bickel, buttinoni, mishra, ohta, peruani, czirok,
stark,weber, angelani,Wan, ghosh, potosky}. The
kinetics of active particles moving in periodic structures could
exhibit peculiar behaviors. Active particles or agents are assumed to have an internal propulsion mechanism, which
may use energy from an external source and transform it under
non-equilibrium conditions into directed motion. Therefore, it would
be significant important to study transport behaviors of active
Brownian particles in entropic potentials. Recently, Ghosh and
coworkers \cite{ghosh}studied the transport of self-propelled
particles in periodic entropic potentials and found that ratcheting
current can be orders of magnitude stronger than for ordinary
thermal potential ratchets. Then, they found that elliptic Janus
particles along narrow two-dimensional channels can show giant
absolute negative mobility \cite{ghosh2} and the mean exit time of
Janus particles in two dimensional cavities is very sensitive
to the cavity geometry, particle shape, and self-propulsion strength
\cite{ghosh3}. Yariv and coworkers \cite{Yariv} studied the transport of Brownian swimmers in a periodically corrugated channel by using the reduced Fokker-Planck approach.

\indent In this paper, we mainly studied the ratchet transport
 of interacting self-propelled particles in periodic entropic potentials. We focus on finding how the interaction between
  active particles affects the entropic transport. From numerical simulations, it is found that upon variation of the interaction strength, the average velocity exhibits  nonmonotonical behaviors. For the attractive case, the average velocity firstly decreases, then increases, and finally decreases to zero.  For the repulsive case,  the interaction affects the transport only for very small interaction strength,  for  the large repulsive strength, the average velocity is independent of the interaction and tends to that in single-particle system. In the regime of small interaction strength, there exists a critical value of the repulsive strength at which the average velocity takes its minimal value, nearly tends to zero.

\section{Model and methods}
\indent In this paper, we consider a set of $N$ interacting self-propelled particles in a periodic
 two-dimensional channel. A self-propelled particle is viewed as characterized by a unit
 vector $\vec{n}_i\equiv (\cos\theta_i,\sin\theta_i)$ in the $xy$ plane, defining the
 direction of the self-propelled velocity. The particles are subjected to both translational
 and rotational diffusion, with coefficients $D_0$ and $D_{\theta}$, respectively.
 The dynamics of the particle $i$ is described by the following overdamped Langevin equations.
\begin{equation}\label{eq1}
\frac{d\vec{r}_i}{dt}=v_0\vec{n}_i-\mu\sum_{j\neq i}\frac{\partial}{\partial \vec{r}_i}V(\vec{r}_i-\vec{r}_j)+\sqrt{2D_0}\vec{\xi}_i(t),
\end{equation}
\begin{equation}\label{eq2}
  \frac{d\theta_i}{dt}=\sqrt{2D_\theta}\xi^{\theta}_i(t),
\end{equation}
where $\mu$ is the mobility and $v_0$ is the magnitude of the self-propelled velocity.  Gaussian white noise terms for both the translational
and rotational motion are characterized by $\langle\vec{\xi}_{i}(t)\rangle = 0$,
 $\langle \xi_{i}^{\alpha}(t)\xi_{j}^{\beta}(s)\rangle = \delta_{ij}\delta_{\alpha\beta}\delta(t-s)$
 and $\langle \xi^{\theta}_i(t)\rangle =0$, $\langle \xi^{\theta}_i(t)\xi^{\theta}_j(s)\rangle = \delta_{ij}\delta(t-s)$, respectively.
 Here $i,j=1,...,N$ label the particles and $\alpha,\beta=x,y$ the coordinates of the space. The symbol $\langle...\rangle$
 denotes an ensemble average over the distribution of the random forces. $\delta$ is the Dirac delta function. For the sake of simplicity, we have ignored hydrodynamic effects.

\begin{figure}[htbp]
\vspace{-4cm}
\begin{center}
\includegraphics[width=16cm]{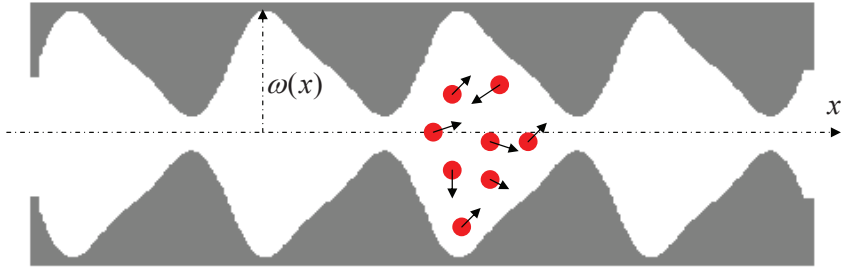}
\vspace{-4cm}
\caption{(Color online) Scheme of the entropic ratchet device: interacting self-propelled particles moving in a two-dimensional channel. The shape is described by the radius $\omega(x)$ of the channel. }\label{1}
\end{center}
\end{figure}

\indent The shape of the channel can be described by its radius $\omega(x)$ shown in Fig. 1
\begin{equation}\label{bd}
    \omega(x)=a[\sin(\frac{2\pi x}{L})+\frac{\Delta}{4}\sin(\frac{4\pi
    x}{L})]+b,
\end{equation}
where $\Delta$ is the asymmetry parameter of the channel shape and $a$ is the parameter that controls the slope of the channel.  The radius at the bottleneck is determined by the parameters $a$, $b$, and $\Delta$.

As for the pair interaction potential $V$, we consider two cases: (A) the attractive potential and (B) the repulsive potential. For case A
\begin{equation}\label{}
  V(r)=\frac{1}{2}k_ar^2,
\end{equation}
and for case B
\begin{equation}\label{}
  V(r)=\frac{k_r}{r},
\end{equation}
where $r$ is the center to center distance between any two particles. $k_a$ and $k_r$ are the attractive and repulsive strength, respectively.

\indent Rectification of Brownian particles has been the focus of a concerted effort, both conceptual and technological, aimed at establishing net particle transport on a periodic substrate in the absence of external biases.  The most important quantity characterizing  the rectification in our system is its directional velocity along $x$ direction. Since the Fick-Jacobs equation corresponding to the Langevin equations (\ref{eq1})and (\ref{eq2}) can not be solved analytically,
we have numerically simulated the overdamped two dimensional dynamics of Brownian particles (\ref{eq1},\ref{eq2}) along with the boundary conditions Eq. (\ref{bd}) using an improved Euler algorithm.   In the asymptotic long-time regime,  the average velocity of particle $i$ along $x$ direction can be obtained from the following formula
           \begin{equation}\label{V}
            v^{\theta_0}_i=\lim_{t\rightarrow\infty}\frac{\langle x_i(t)\rangle_{\theta_0}}{t},
            \end{equation}
where $\theta_0$ is the initial angle of the trajectory. The average velocity after a second average over all $\theta_0$ is
\begin{equation}\label{}
  v_i=\frac{1}{2\pi}\int_0^{2\pi}d\theta_0 v^{\theta_0}_i.
\end{equation}
The full  average velocity is $\overline{v}=\frac{\sum_{i=1}^{N}v_i}{N}$. For the convenience of discussion, we define the scaled average velocity $v_s=\overline{v}/v_0$ through the paper.

\section{Results and Discussion}
\indent In our simulations, the integration step time $\Delta t$ was chosen to be smaller than $10^{-4}$ and the total integration
time was more than $3\times 10^5$. The stochastic averages reported above were obtained as ensemble averages over $3 \times 10^{4}$
trajectories with random initial conditions. Unless otherwise noted, our simulations are under the parameter sets: $a=\frac{1}{2\pi}$, $b=\frac{1.2}{2\pi}$, $\Delta=1.0$, and $N=4$. The simulation results are reported in Figs. 2-6.
\begin{figure}[htbp]
\begin{center}\includegraphics[width=10cm]{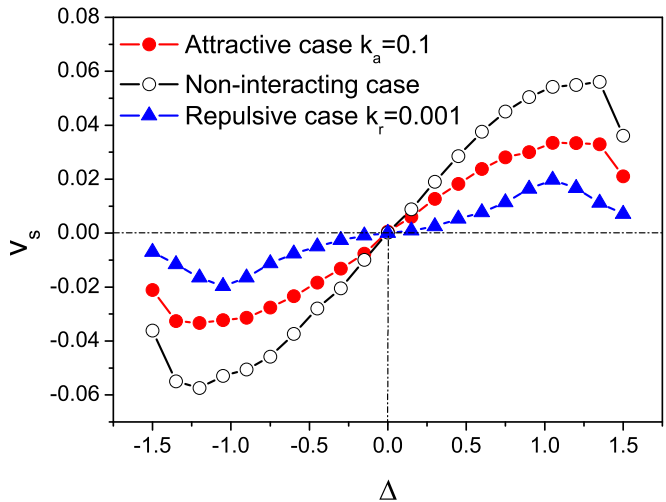}
\caption{(Color online) Average velocity $v_s$ versus the asymmetric parameter $\Delta$ for three cases. The other parameters are $v_0=5.0$, $k_a=0.1$, $k_r=0.001$, $D_0=0.1$, and $D_{\theta}=0.1$.}\label{1}
\end{center}
\end{figure}

\indent In Figure 2, we plot the average velocity $v_s$ as a function of the asymmetric parameter $\Delta$ for three cases.  It is found that the direction of the transport is completely determined by the symmetry  of the channel.  The average velocity is positive for $\Delta>0$,  zero at $\Delta=0$, and negative for $\Delta<0$.  When $|\Delta|\rightarrow 0$,  the channel is symmetric, so there is no net current.  When $|\Delta|>\Delta_c$, the channel is blocked, no particles can pass across the cell of the channel. Therefore, there exists an optimal value of $|\Delta|$ at which the average velocity takes its maximal value.

\begin{figure}[htbp]
\begin{center}\includegraphics[width=10cm]{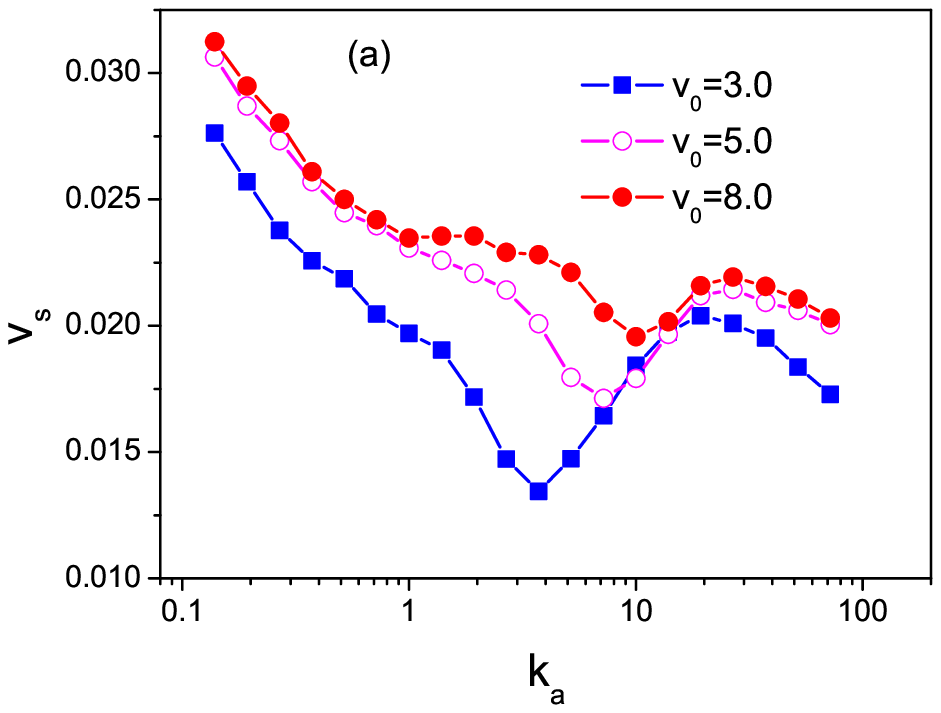}
\includegraphics[width=10cm]{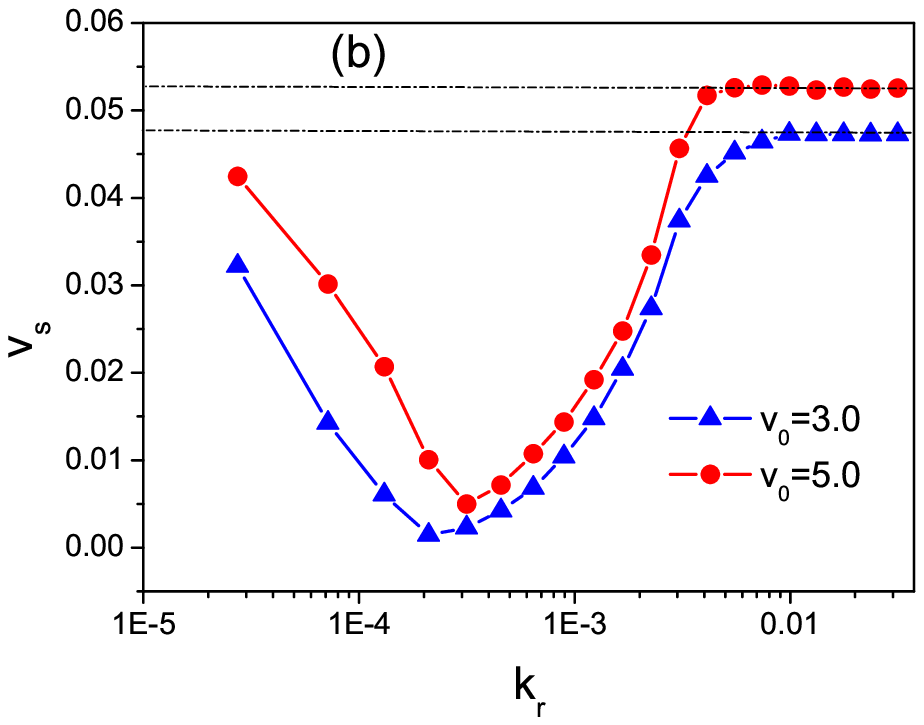}
\caption{(Color online) Average velocity $v_s$ versus the strength of the interaction potential. (a) For attractive particles ($k_a$). (b) For repulsive particles ($k_r$). The other parameters are $v_0=5.0$, $D_{\theta}=0.1$, and $D_0=0.1$}\label{1}
\end{center}
\end{figure}

\indent Figure 3(a) shows the average velocity $v_s$ as a function
of the attractive strength $k_a$ for different values of $v_0$. On
increasing $k_a$ from zero, the average velocity $v_s$ firstly decreases to
its minimal value, then increases, and finally decreases to zero.
There exist a valley and a peak in the curve and the
average velocity takes its minimal value when $k_a\simeq v_0$.  This
features can be explained by the following considerations. The
attractive interaction in the system can cause two results: (A)
reducing the self-propelled driving,  which blocks the ratchet
transport and (B) activating motion in analogy with thermal noise
activated motion for a single stochastically driven ratchet, which
facilitates the ratchet transport. When $k_a\rightarrow 0$, the
average velocity attains a constant value (average velocity in
noninteracting case).  When $k_a<v_0$, the factor A dominates the
transport, so the average velocity $v_s $ decreases when $k_a$
increases.  When $k_a>v_0$, the factor B gradually becomes
significant and the average velocity increases with $k_a$.  However,
for very large values of $k_a$ ($k_a\rightarrow \infty$)
 all particles are gathered as a single particle, both many-body effects and
 the individual self-propelled driving can be neglected, so the average velocity tends to zero.
 The average velocity of interacting attractive  particles is always smaller than that for the
 noninteracting case, which indicates that the attractive interaction always blocks the rectification.
 However, for passive Brownian particles, there exist some values of the attractive strength where the interaction
 can facilitate the rectification \cite{Csahok}.

 \indent Figure 3 (b) describes the average velocity $v_s$ as a
 function of the repulsive strength $k_r$ for different values of
 $v_0$. It is found that on increasing $k_r$, the average velocity
 $v_s$ first decreases to nearly zero value, then increases, and
 finally tends to a constant. The repulsive interaction in the system can also cause two results:
 (A) reducing the self-propelled driving (blocking the ratchet transport) and
 (B) dispersing Brownian particles (facilitating the directed transport).  When $k_r$ is very small,
 the factor A dominates the transport and the dispersing effect
 can be neglected.  Therefore, the average velocity $v_s$ decreases when $k_r$ increases from zero.
 However, when $k_r$ becomes large, the dispersing effect becomes significant, so the average velocity $v_s$ increases.
 For large values of $k_r$, the dispersing effect completely dominates the transport and the interaction between Brownian particles
 can be ignored.  When $k_r\rightarrow 0$ or $ k_r  \rightarrow\infty $,  the average velocity $v_s$  tends to
 the average velocity in the single-particle system, therefore, there exits a critical value of $k_r$ (very small value)  at which
 the average velocity $v_s$ is minimal, and nearly tends to zero.   The critical value of $k_r$ depends on the parameters of the system.

\begin{figure}[htbp]
\begin{center}\includegraphics[width=10cm]{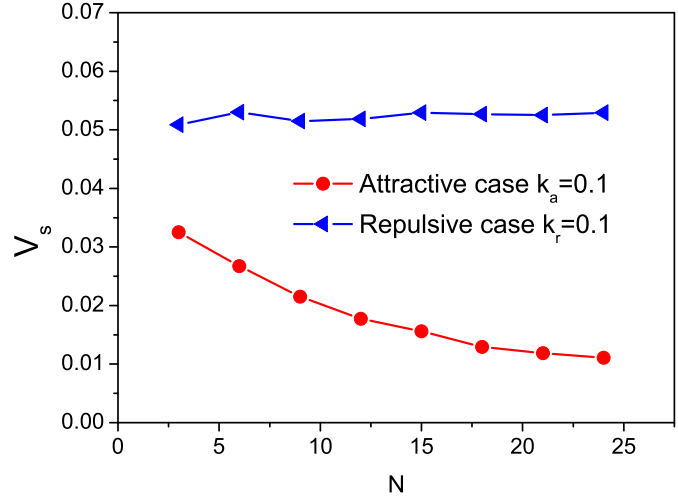}
\caption{(Color online) Average velocity $v_s$  versus particle number $N$ for both attractive and repulsive particles. The other parameters are $v_0=5.0$, $D_{\theta}=0.1$, $k_a=0.1$, $k_r=0.1$, and $D_0=0.1$}\label{1}
\end{center}
\end{figure}
\indent Figure 4 shows the dependence of the average velocity $v_s$ on the particle number $N$ for both attractive and repulsive cases.  For the attractive case,  the average velocity $v_s$ decreases monotonically with increasing $N$.  The average velocity will tend to zero when $N \rightarrow \infty$.  The is because the effective attractive driving for large number  $N$ can be neglected after the average .  For the repulsive case, the average velocity $v_s$ is independent of the particle number $N$ when $k_r>0.01$.  In the infinitely long channel, the repulsive forces between particles disperse particles,  the distance between particles become longer and the repulsive  forces gradually disappear.

\begin{figure}[htbp]
\begin{center}\includegraphics[width=10cm]{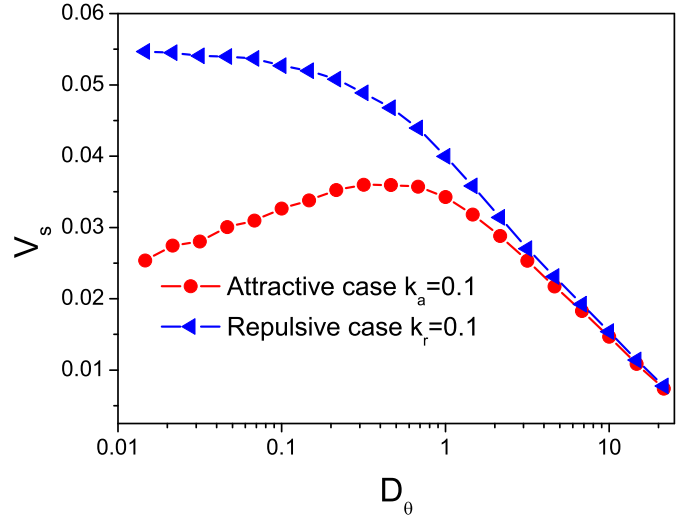}
\caption{(Color online) Average velocity $v_s$ versus the rotational diffusion $D_{\theta}$ for both attractive and repulsive particles. The other parameters are  $v_0=5.0$, $k_a=0.1$, $k_r=0.1$, and $D_0=0.1$.}\label{1}
\end{center}
\end{figure}
\indent In Fig. 5, we explore the average velocity $v_s$ as a function of the rotational diffusion $D_\theta$ for both attractive and repulsive cases.   When $D_\theta\rightarrow \infty$, transport behaviors are similar for two cases.  The self-propelled velocity changes its direction very fast.  The self-propelled velocity acts as a zero mean white noise and the nonequilibrium driving in the system disappears, so no directed transport occurs and the average velocity tends to zero.  For small values of $D_\theta$, the transport behaviors are different for two cases:  (A) For the repulsive case ($k_r=0.1$),  the average velocity decreases monotonically when $D_\theta $ increases.  Especially,  when $D_\theta  \rightarrow 0$, the average velocity tends to a  saturate value.  In the adiabatic limit, the repulsive force can be expressed by two opposite static force $v_0$ and $-v_0$, yielding the mean zero velocity $v_s=[v_s(v_0)+v_s(-v_0)]/2$, which  is similar to  the  single-particle thermal ratchet\cite{ratchet}.  (B) For the attractive case,  there exists an optimal value of $D_\theta$ at which the average velocity $v_s$ is maximal.  The increase of $D_\theta$ in this case can  reduce the self-propelled driving(blocking the ratchet transport) and activate the attractive Brownian particles(facilitating the directed transport).   When $D_\theta$ increases from zero,  the latter factor dominates the transport and the average velocity $v_s$ increases to its maximal value.  For further increasing $D_\theta$, the former factor takes effect, the average velocity $v_s$ decreases.

\begin{figure}[htbp]
\begin{center}\includegraphics[width=10cm]{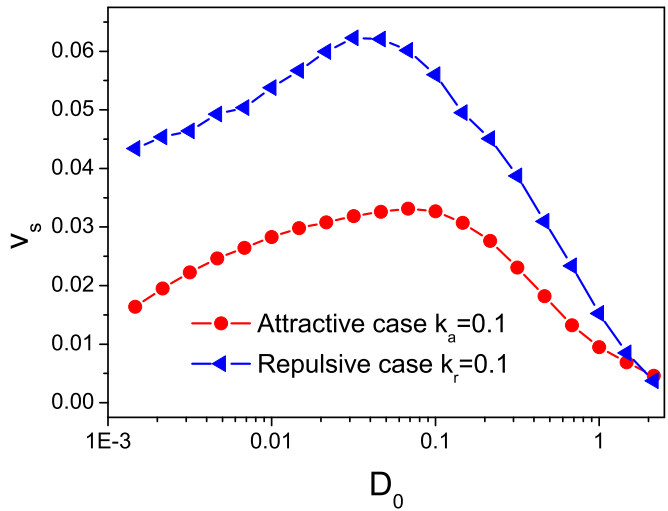}
\caption{(Color online) Average velocity $v_s$ versus the translational diffusion $D_{0}$ for both attractive and repulsive particles. The other parameters are $v_0=5.0$, $k_a=k_r=0.1$, and $D_{\theta}=0.1$. }\label{1}
\end{center}
\end{figure}
\indent Figure 6 illustrates the impact of the translational diffusion $D_0$  on the average velocity $v_s$ for both the attractive and repulsive cases.  From the figure, we can see that the curves are similar for two cases.  When $D_0\rightarrow 0$, the particles will stay at the bottom of the channel  and cannot pass the entropic barrier, so the average velocity $v_s$ goes to zero. When $D_0\rightarrow \infty$, the translational diffusion is very large, the effect of the asymmetric entropic barrier disappears and the average velocity $v_s$ tends to zero. Therefore, there exists an optimal $D_0$ value where the average velocity $v_s$ is maximal.

\section {Concluding remarks}
\indent In this paper,  we numerically studied the directed transport of interacting self-propelled Brownian particles in a two-dimensional  periodic channel. It is found that the self-propelled velocity acts as the nonequilibrium driving,  which can break the themodynamical equilibrium and induce the directed transport. The direction of the transport is completely determined by the symmetry of the channel shape.  The interaction between Brownian particles can significantly affect the directed transport.
 For the attractive case: (1) on increasing the strength $k_a$ from zero, the average velocity $v_s$ first decreases to
its minimal value, then increases, and finally decreases to zero; (2) the average velocity $v_s$ decreases monotonically with increase of  the particle number $N$;  (3) there exists an optimal value of $D_\theta$ at which the average velocity $v_s$ takes is maximal value, which is different from the noninteracting case, where the average velocity decreases  monotonically with increase of $k_r$.   For the repulsive case: (1) the interaction affects the transport only for very small interaction strength,  for  the large repulsive strength, the average velocity is independent of the interaction and tends to that in single-particle system. In the regime of small interaction strength, there exists a critical value of the repulsive strength at which the average velocity takes its minimal value, nearly tends to zero; (2) the average velocity $v_s$ is independent of the particle number $N$ when $k_r>0.01$; (3) the average velocity $v_s$ decreases monotonically with increase of the rotational diffusion $D_\theta$, especially, it tends to a saturate value when $D_\theta\rightarrow 0$.  In addition, the average velocity of interacting active Brownian particles is always less than that in the noninteracting case (which indicates the interaction always cannot facilitate the rectification), which is different from the passive case, where the interaction may facilitate the rectification.

\indent  This work was supported in part by the National Natural Science Foundation
of China (Grant Nos. 11175067, 11004082, and 11205044), the PCSIRT (Grant No. IRT1243), the Natural
Science Foundation of Guangdong Province, China (Grant No. S2011010003323).

\end{document}